# The mean first passage time in energy-diffusion controlled regime with power-law distributions

Zhou Yanjun and Du Jiulin

*Department of Physics, School of Science, Tianjin University, Tianjin 300072, China*

**Abstract:** Based on the mean first passage time (MFPT) theory, we derive an expression of the MFPT in the energy-diffusion controlled regime with a power-law distribution. We discuss the finite barrier effect (i.e. thermal energy $k_BT$ is not small with respect to the potential barrier $E_b$) and compare it with the Kramers infinite barrier result both in a power-law distribution and in a Maxwell-Boltzmann distribution. It is shown that the MFPT with the power-law distribution extends Kramers' low damping result to a relatively low barrier. We pay attention to the energy-diffusion controlled regime, which is of great interest in the context of Josephson junctions, and study how the power-law parameter $\kappa$ affects the current distribution function in experiment with Josephson junctions.

**Keywords**：Mean first passage time; Power-law distribution; Energy-diffusion

## 1. Introduction

In 1940, Kramers proposed the thermal escape of a Brownian particle from a metastable well, and, according to the very low and intermediate to high dissipative coupling to the bath, he presented three explicit formulae for the escape rates in the low damping, intermediate-to-high damping and very high damping respectively, all of which have received much attention and interest in physics, chemical reaction rate, biology [1-3], among others. Thermal activation has already received a lot of attention in the past. However, most of the work refers to the thermal equilibrium case, while in view of many applications a correct understanding of the nonequilibrium case is necessary especially in low damping systems. Because the coupling to the bath is very weak and the time to reach thermal equilibrium very long in low damping systems, the particles may escape before thermal equilibrium is established and thus nonequilibrium effects dominate the process [4]. Therefore, in this paper, starting from the mean first passage time (MFPT) theory, we develop a solution to the problem of thermal activation in nonequilibrium conditions.

Recently, plenty of observations in experiments have shown that non-Maxwell-Boltzmann (MB) distributions and power-law distributions are quite common in nonequilibrium complex systems, such as glasses [5], disordered media [6-8], folding of proteins [9], single-molecule conformational dynamics [10, 11], trapped ion reactions [12], chemical kinetics, and biological and ecological population dynamics [13, 14], reaction–diffusion processes [15], chemical reactions [16], combustion processes [17], gene expression [18], cell reproductions [19], complex cellular networks [20], small organic molecules [21], astrophysical and spatial plasmas [22], and others. The typical forms of such power-law distributions have included the kappa-distributions in the solar wind and interplanetary space plasmas [22-26], the *q*-distributions in complex systems within nonextensive statistical mechanics [27], and those *α*-distributions noted widely in physics, chemistry, biology and elsewhere such as $P(E) \sim E^{-\alpha}$, with an index $\alpha > 0$ [12, 15, 16, 21, 23, 28]. These power-law distributions may lead to processes different from those in the realm governed by Boltzmann–Gibbs statistics with MB distribution. Simultaneously, a class of statistical mechanical theories studying the power-law distributions in complex systems has been constructed, for instance, by generalizing Boltzmann entropy to Tsallis entropy [27], by generalizing Gibbsian theory [29] to a system away from thermal equilibrium, and so forth. The stochastic dynamical theory of power-law distributions has been developed by means of studying the Brownian motion in a complex system [30-32], which has led to a generalized fluctuation-dissipation relation (FDR), a generalized Smoluchowski equation and a generalized Klein- Kramers equation.

Non-MB distributions and power-law distributions might be associated with anomalous transports. Some examples for anomalous diffusions have been reported, such as that in living cells [33], and that in viscoelastic environments [34,35]. Anomalous diffusions have been studied sometimes by, for example, a fractional Langevin equation [36], a fractional Fokker-Planck equation [37], and the fractional kinetics [38].

One of the most important thermal escape theories is the first passage time theory [39, 40]. Suppose that the motions of Brownian particles are bounded in a finite space *V* with an absorbing boundary $\Sigma$, and the position and momentum are *x* and *p*, respectively. The motions are governed by the Langevin equation [30],

$$\frac{dx}{dt} = \frac{p}{m}, \qquad \frac{dp}{dt} = -\frac{dU(x)}{dx} - \gamma p + \eta(x, p, t), \qquad (1)$$

where *m* is the mass of the particle, *U*(*x*) is a potential field, *γ* is the friction coefficient. Usually, in normal conditions, the friction coefficient is regarded as a constant. But when the Brownian particle moves in an inhomogeneous complex medium, it may be considered as a function of the variables *x* and *p*, i.e. *γ*=*γ*(*x,p*). For a complex system, the noise may be considered as inhomogeneous in (*x, p*) and so it is also a function of the variables, i.e. *η* =*η*(*x,p,t*) is a multiplicative (space/velocity dependent) noise [3]. As usual, it is assumed that the noise is Gaussian, with zero average and delta-correlated in time *t*, such that it satisfies,

$$\langle \eta(x,p,t) \rangle = 0, \quad \langle \eta(x,p,t)\eta(x,p,t') \rangle = 2D(x,p)\delta(t-t'). \qquad (2)$$

Thus, the correlation strength of multiplicative noise (i.e. diffusion coefficient) *D*(*x, p*) is a function of the variables *x* and *p*.

If *ρ*(*x,p,t*) is the probability distribution of the particles that have not left *V* by time *t*, then it satisfies the Fokker-Planck equation,

$$\frac{\partial \rho}{\partial t} = -\frac{p}{m}\frac{\partial \rho}{\partial x} + \frac{\partial}{\partial p}\left(\frac{dU}{dx} + \gamma p\right)\rho + \frac{\partial}{\partial p}\left(D\frac{\partial \rho}{\partial p}\right), \quad (3)$$

where the initial condition and the absorbing boundary condition are $\rho(x,p,0) = \delta(x-x_0)\delta(p-p_0)$ and $\rho(x,p,t)|_\Sigma = 0$, respectively. In an inhomogeneous complex media, if the diffusion coefficient and the friction coefficient satisfy the generalized fluctuation-dissipation relation [30],

$$D = m\gamma\beta^{-1}(1 - \kappa\beta E), \quad (4)$$

where $E \equiv p^2/2m + U(x)$ is the energy, $\beta = 1/k_B T$, $k_B$ is the Boltzmann constant, $T$ is the temperature, $\kappa$ is a parameter, and $\kappa \neq 0$ measures a distance away from thermal equilibrium, then substituting Eq. (4) into Eq. (3), one can show that the stationary-state solution becomes the power-law $\kappa$-distribution [30],

$$\rho_s(E) = Z_\kappa^{-1}(1 - \kappa\beta E)_+^{-1/\kappa}, \quad (5)$$

where the normalization constant is $Z_\kappa = \iint dx dp (1 - \kappa\beta E)_+^{\frac{1}{\kappa}}$. It is clear that the generalized FDR is a condition under which the power-law $\kappa$-distribution can be created from the stochastic dynamics of the Langevin equations.

We introduce the Fokker-Planck operator and its adjoint operator [39],

$$f = -\frac{p}{m}\frac{\partial}{\partial x} + \frac{\partial}{\partial p}\left(\frac{dU}{dx} + \gamma p\right) + \frac{\partial}{\partial p}D\frac{\partial}{\partial p}, \quad (6)$$

$$f^\dagger = \frac{p}{m}\frac{\partial}{\partial x} - \left(\frac{dU}{dx} + \gamma p\right)\frac{\partial}{\partial p} + \frac{\partial}{\partial p}D\frac{\partial}{\partial p}, \quad (7)$$

and as an initial value problem the operator solution is $\rho(x,p,t) = e^{tf}\delta(x-x_0)\delta(p-p_0)$. Integrating $\rho(x,p,t)$ over the volume $V$, the numbers of all starting points $s(t, x_0, p_0)$ that are still in $V$ at time $t$ are obtained, which depends on the initial values $x_0$ and $p_0$. The number difference of the initial points that have not left before time $t$ but have left during the time interval $dt$ after time $t$ determines the probability density, $p(t, x_0, p_0)$, of lifetime,

$$p(t, x_0, p_0) = -\frac{ds(t, x_0, p_0)}{dt}. \quad (8)$$

Because of the noise, even the same initial positions will lead to different first passage time; hence, the mean first passage time $\tau(x, p)$ is introduced. There is a more direct way to calculate MFPT. Namely, by using the adjoint operator $f^\dagger$ to operate on the MFPT, the MFPT is determined by solving the inhomogeneous adjoint equation [39],

$$\frac{p}{m}\frac{\partial \tau(x,p)}{\partial x}-\left(\frac{dU}{dx}+\gamma p\right)\frac{\partial \tau(x,p)}{\partial p}+\frac{\partial}{\partial p}\left(D\frac{\partial}{\partial p}\tau(x,p)\right)=-1, \tag{9}$$

and $\tau(x,p)=0$ on $\Sigma$. Eq. (9) is the starting point of the following work.

This paper is organized as followed. In section 2, Eq. (9) is transformed into an energy-diffusion equation satisfied by the MFPT. In section 3, we solve the equation and get an expression for the MFPT in the normal FDR ($\kappa=0$) and the generalized FDR ($\kappa\neq 0$), respectively. Then we discuss the finite barrier effects by comparing the general result with Kramers infinite barrier result (namely, $\beta E_b \gg 1$). In the case of Josephson junctions [4, 41], for instance, where the barrier height is $\beta E_b = \frac{\gamma_0}{2}[-\pi\alpha+2(\alpha\sin^{-1}\alpha+\sqrt{1-\alpha^2})]$, where $\gamma_0 = \Phi_0 I_C/\pi k_B T$, $\alpha \equiv I_0/I_C$, $\Phi_0$ is the magnetic flux quantum, $I_0$ is the bias current of the circuit and $I_C$ is the critical current, we see that the barrier decreases with the increasing of $\alpha$, and that $\beta E_b$=0 when $\alpha=1$. Another example is the single-molecular pulling experiment [42, 43]; the barrier height is $\Delta G$ in the absence of the external force $F$. When $F$ is applied to the experiment, the barrier height $\beta E_b$ decreases with increasing $F$. Therefore, research on the finite barrier effect is important. In section 4, we take Josephson junctions as an example to study how the power-law parameter $\kappa$ affects the current distribution function in experiment with Josephson junctions. Finally in section 5, the conclusion is given.

## 2. The energy-diffusion equation for the MFPT

For a small damping, Brownian force causes only a tiny perturbation in the undamped energy during the oscillation in the well; therefore the energy is a slowly varying quantity but the phase is a quickly varying quantity. One may write the Fokker-Planck equation in the canonical variables ($x$, $p$) as a diffusion equation in the energy ($E$) and phase ($w$), and make an average for the density over the fast phase variable, and then get an energy diffusion equation in the slow (almost conserved) energy variable [3]. The time average along a trajectory is defined [3] as

$$\overline{\rho}(E,t) \equiv \frac{1}{T'}\int_0^{T'} \rho(E,w,t)dw, \quad dw=dt,$$

where $T'$ is the time required to execute one cycle of almost one periodic motion. The chain rule is used below to transform Eq. (9) into a function of both the energy $E$ and the phase $w$,

$$\frac{\partial}{\partial x}=\frac{dU}{dx}\frac{\partial}{\partial E}+\frac{\partial w}{\partial x}\frac{\partial}{\partial w}=\frac{dU}{dx}\frac{\partial}{\partial E}+\frac{\partial w}{\partial t}\frac{\partial t}{\partial x}\frac{\partial}{\partial w}=\frac{dU}{dx}\frac{\partial}{\partial E}+\frac{m}{p}\frac{\partial}{\partial w}, \quad \frac{\partial}{\partial p}=\frac{p}{m}\frac{\partial}{\partial E}. \tag{10}$$

Eq. (9) is then rewritten as

$$\frac{\partial \tau(E,w)}{\partial w} - \gamma \frac{p^2}{m}\frac{\partial \tau(E,w)}{\partial E} + \frac{p^2}{m^2}\frac{\partial D}{\partial E}\frac{\partial \tau(E,w)}{\partial E} + \frac{D}{m}\frac{\partial \tau(E,w)}{\partial E} + \frac{Dp^2}{m^2}\frac{\partial^2 \tau(E,w)}{\partial E^2} = -1. \qquad (11)$$

Making an average over the phase $w$ of Eq. (11), one has

$$\overline{\frac{\partial \tau(E,w)}{\partial w}} - \overline{\gamma}\frac{\overline{p^2}}{m}\overline{\frac{\partial \tau(E,w)}{\partial E}} + \frac{\overline{p^2}}{m^2}\overline{\frac{\partial D}{\partial E}}\overline{\frac{\partial \tau(E,w)}{\partial E}} + \frac{\overline{D}}{m}\overline{\frac{\partial \tau(E,w)}{\partial E}} + \overline{D}\frac{\overline{p^2}}{m^2}\overline{\frac{\partial^2 \tau(E,w)}{\partial E^2}} = -1, \qquad (12)$$

and using the definition of time average, the first term on the left-hande side of Eq.(12) becomes

$$\overline{\frac{\partial \tau(E,w)}{\partial w}} \equiv \frac{1}{T'}\int_0^{T'} \frac{\partial \tau(E,w)}{\partial w} dw \approx \frac{1}{T'}\int_0^{T'} d\tau(E,w) = 0, \qquad (13)$$

where the integral is taken over one complete cycle of the motion, thus one has $\tau(E,0) = \tau(E,T')$ and the integral is zero. Note that Eq. (13) holds only approximately for slow varying of $E$, since $d\tau(E,w) = \frac{\partial \tau}{\partial E}dE + \frac{\partial \tau}{\partial w}dw \approx \frac{\partial \tau}{\partial w}dw$ [3]. The kinetic energy term

$$\overline{p^2} = \frac{1}{T'}\int_0^{T'} p^2 dw = \frac{\omega}{2\pi}\int_0^{2\pi/\omega} p^2 dt = \frac{\omega}{2\pi}\int_0^{2\pi/\omega} pm\dot{x}dt = \frac{m\omega}{2\pi}\oint pdx = \frac{m\omega I}{2\pi}, \qquad (14)$$

is calculated by taking the periodic time in the well as $T' = 2\pi/\omega(E)$, where $\omega(E)$ is an angular frequency of the oscillation and $I$ is an action over the period defined [3] by $I(E) = \oint_{E=Const} pdx$. Thus, Eq. (12) becomes

$$-\overline{\gamma}\frac{\omega I}{2\pi}\overline{\frac{\partial \tau(E,w)}{\partial E}} + \frac{\omega I}{2\pi m}\overline{\frac{\partial D}{\partial E}}\overline{\frac{\partial \tau(E,w)}{\partial E}} + \frac{\overline{D}}{m}\overline{\frac{\partial \tau(E,w)}{\partial E}} + \overline{D}\frac{\omega I}{2\pi m}\overline{\frac{\partial^2 \tau(E,w)}{\partial E^2}} = -1. \qquad (15)$$

The Hamiltonian of the motion is $E = p^2/2m + U(x)$, and it can be turned into the differential equation,

$$\frac{dx}{dt} = \pm\sqrt{2[E-U(x)]/m}. \qquad (16)$$

Taking the positive sign and integrating this differential equation between 0 and $a$ yields

$$\int_0^a \frac{dx}{\sqrt{2[E-U(x)]/m}} = t. \qquad (17)$$

Eq. (17) determines the angular frequency of the oscillation over one period,

$$\oint \frac{dx}{\sqrt{2[E-U(x)]/m}} = 2\pi/\omega(E). \qquad (18)$$

Denoting that

$$I' = \oint \frac{dx}{\sqrt{2[E-U(x)]/m}}, \quad (19)$$

Eq. (15) is finally written as

$$\left(\frac{I}{mI'}\frac{\partial \overline{D}}{\partial E} - \overline{\gamma}\frac{I}{I'} + \frac{\overline{D}}{m}\right)\frac{\partial \overline{\tau(E,w)}}{\partial E} + \overline{D}\frac{I}{mI'}\frac{\partial^2 \overline{\tau(E,w)}}{\partial E^2} = -1. \quad (20)$$

If we drop the superscript "−" in Eq. (20), it then becomes

$$\left(\frac{I}{mI'}\frac{\partial D}{\partial E} - \gamma(E)\frac{I}{I'} + \frac{D}{m}\right)\frac{\partial \tau(E)}{\partial E} + D(E)\frac{I}{mI'}\frac{\partial^2 \tau(E)}{\partial E^2} = -1, \quad (21)$$

Eq. (21) is a general energy-diffusion equation of MFPT, one key result of this paper. In section 3, for further study, two cases will be considered.

## 3. MFPT for the normal FDR ($\kappa = 0$) and the generalized FDR ($\kappa \neq 0$)

3.1. The case when $\kappa = 0$:

It is well known that the stochastic dynamics to generate a MB distribution for a thermal equilibrium system can be well explained by the Langevin equations, which model the motion of a Brownian particle moving in a potential field and in the medium with a constant friction coefficient and white noise. And the diffusion constant $D$ is related to the friction constant $\gamma$ by the normal FDR, $D = m\gamma\beta^{-1}$, thus Eq. (21) becomes

$$\left(-\gamma\frac{I}{I'} + \gamma\beta^{-1}\right)\frac{\partial \tau(E)}{\partial E} + \gamma\beta^{-1}\frac{I}{I'}\frac{\partial^2 \tau(E)}{\partial E^2} = -1. \quad (22)$$

Supposing that the initial energy $E$ is $0 < E < E_b$, $E_b$ is the energy at the barrier, and the absorbing boundary condition is $\tau(E_b) = 0$, the solution of Eq. (22) is

$$\tau(E) = \int_E^{E_b} \frac{\beta I'}{I\gamma} e^{\beta E'} dE' \int_0^{E'} dE'' e^{-\beta E''}. \quad (23)$$

If the angular frequency of oscillation in the well can be approximately taken as constant, $\omega(E) \approx \omega_0$, then $\omega(E) = 2\pi dE/dI$ becomes $I\omega_0/2\pi = E$, and $I'/I = 1/E$. Eq. (23) becomes

$$\tau(E) = \int_E^{E_b} \frac{\beta}{E\gamma} e^{\beta E'} dE' \int_0^{E'} dE'' e^{-\beta E''}, \quad (24)$$

If the thermal energy $k_BT$ is much smaller than the potential barrier $E_b$, i.e. $\beta E_b \gg 1$, the second integral is dominated by small $E''$, and the first integral is dominated by $E'$ near $E_b$. Therefore, the MFPT is

$$\tau = \frac{e^{\beta E_b}}{\gamma \beta E_b}, \qquad (25)$$

which is the well-known Kramers expression for the case of low damping, so Eq. (24) is a general expression of the MFPT for an arbitrary $\beta E_b$.

Equivalently, Eq. (24) can be expressed as

$$\tau(x) = \frac{1}{\gamma} \int_x^{\beta E_b} \frac{1}{y} e^y dy \int_0^y dz e^{-z} \qquad (26)$$

if we make a substitution, $\beta E = x$. In Fig.1, we show the results of a numerical integral over Eq. (26) and compared the result for arbitrary $\beta E_b$ with the Kramers result for $\beta E_b \gg 1$. For convenience, it was assumed that the initial energy is zero.

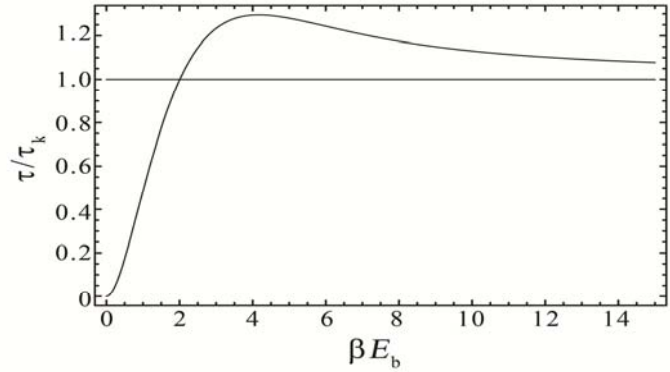

Fig.1 Theoretical estimation of the MFPT normalized by the Kramers result for low damping for arbitrary $\beta E_b$; $\tau/\tau_K = 1$ is also plotted.

Fig.1 shows a quantitative estimation of the accuracy of Kramers result for low damping for $\beta E_b$. The ratio $\tau/\tau_K$ for small $\beta E_b$ is different greatly from that for large $\beta E_b$. The maximal value of the ratio is $\tau/\tau_K = 1.30$ at $\beta E_b = 4.17$, and then the ratio reduces gradually. For instance, $\tau/\tau_K = 1.13$ at $\beta E_b = 10$ and $\tau/\tau_K = 1.09$ at $\beta E_b = 13$. As can be seen, the finite barrier effect is not very severe (below 10%) and the Kramers low damping result is approximately accurate only if $\beta E_b$ is above 13. However, we should consider the finite effect if $\beta E_b$ is below this value.

3.2. The case when $\kappa \neq 0$

We know that when a system reaches its thermal equilibrium state, there exists a standard FDR and the system follows a MB distribution. We mentioned in section 1 that a system in low damping is not in equilibrium, and this simple FDR needs to be generalized. The generalized FDR was derived in Ref.[30] for a system with power-law distributions, i.e. $D(E) = m\gamma(E)\beta^{-1}(1-\kappa\beta E)$, where the diffusion coefficient and the friction coefficient can be both a function of the energy. Therefore, Eq. (21) becomes

$$\left(\frac{I}{I'}\beta^{-1}\frac{\partial\gamma}{\partial E}(1-\kappa\beta E)-(\kappa+1)\gamma\frac{I}{I'}+\gamma\beta^{-1}(1-\kappa\beta E)\right)\frac{\partial\tau(E)}{\partial E}$$

$$+\gamma\beta^{-1}(1-\kappa\beta E)\frac{I}{I'}\frac{\partial^2\tau(E)}{\partial E^2}=-1, \tag{27}$$

and its solution is

$$\tau(E)=\int_E^{E_b}\frac{\beta}{E'}\frac{dE'}{\gamma(E')(1-\kappa\beta E')^{(1+\kappa)/\kappa}}\int_0^{E'}dE''(1-\kappa\beta E'')^{1/\kappa}. \tag{28}$$

The following two cases are taken into consideration:

(I). If the friction coefficient $\gamma(E)$ is a constant, then Eq. (28) becomes

$$\tau(E)=\frac{\beta}{\gamma}\int_E^{E_b}\frac{dE'}{E'(1-\kappa\beta E')^{(1+\kappa)/\kappa}}\int_0^{E'}dE''(1-\kappa\beta E'')^{1/\kappa}. \tag{29}$$

Making the same substitution as in section 3.1, Eq. (29) becomes

$$\tau(x)=\frac{1}{\gamma}\int_x^{\beta E_b}\frac{dy}{y(1-\kappa y)^{(1+\kappa)/\kappa}}\int_0^y dz(1-\kappa z)^{1/\kappa}. \tag{30}$$

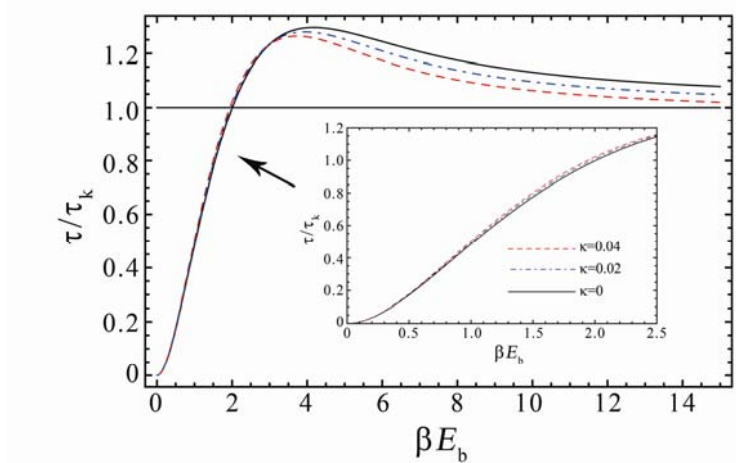

Fig.2 Theoretical estimation of the MFPT normalized by the Kramers low damping result with power-law distributions for the arbitrary $\beta E_b$ and different power-law parameters, the inset corresponds to the low $\beta E_b$.

In Fig.2, following the same way as in section 3.1, we plot a general result normalized by the Kramers result for low damping for an arbitrary $\beta E_b$. The Kramers MFPT result for low damping with the power-law distribution is given in the appendix (see Eq. (A.12)). Here we should pay attention to the cutoff condition for the power-law distribution if $\kappa>0$. If $\kappa>0$, there is a cutoff condition so as to keep $(1-\kappa y)^{1/\kappa}$ being physical. It was shown that if $\beta E_b$ is very small, the curves for the ratio $\tau/\tau_K$ with three different values of the $\kappa$-parameter almost overlapped. Because there is almost no difference between the function $e^x$ and the function $e_\kappa^x = (1-\kappa x)^{1/\kappa}$ if $\beta E_b$ is small, the ratio $\tau/\tau_K$ depends only on $\beta E_b$. With $\beta E_b$ increasing, $\tau/\tau_K$ becomes larger, and at $\beta E_b = 3.74$ it reaches its maximum $\tau/\tau_K = 1.26$, being smaller than that for the case when $\kappa = 0$ (i.e. $\tau/\tau_K = 1.30$ and $\beta E_b = 3.74$). For instance, $\tau/\tau_K = 1.10$ at $\beta E_b = 8$ and $\tau/\tau_K = 1.06$ at $\beta E_b = 10$, all of which are much closer to 1 than that for $\kappa = 0$ case. It is shown that, within an accuracy of 10%, the Kramers low damping result for the power-law distribution is approximately accurate for $\beta E_b$ above 8 in the case of $\kappa \ne 0$, compared with $\beta E_b$ above 13 in the normal case, $\kappa = 0$. Thus, the Kramers low damping result for a power-law distribution has a wider application in the range of relatively low $\beta E_b$.

(II). If the diffusion coefficient $D(E)$ is a constant, then Eq. (28) becomes,

$$\tau(E) = \frac{m}{D} \int_E^{E_b} \frac{dE'}{E'(1-\kappa\beta E')^{1/\kappa}} \int_0^{E'} dE''(1-\kappa\beta E'')^{1/\kappa} . \qquad (31)$$

Making the substitution of $\beta E = x$, Eq. (31) becomes

$$\tau(x) = \frac{m}{D\beta} \int_x^{\beta E_b} \frac{dy}{y(1-\kappa y)^{1/\kappa}} \int_0^y dz (1-\kappa z)^{1/\kappa} . \qquad (32)$$

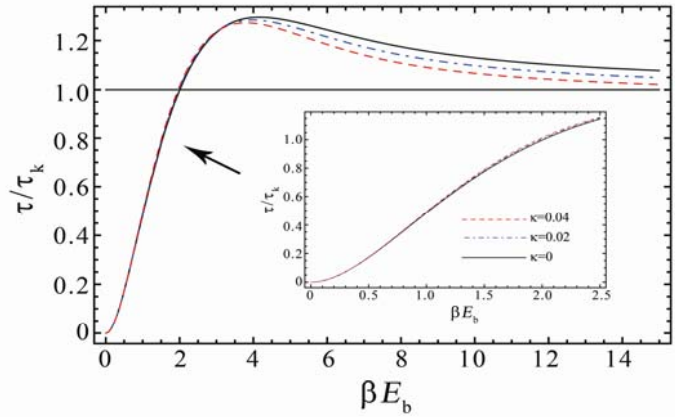

Fig.3 Theoretical estimation of the MFPT normalized by the Kramers low damping result with power-law distributions for the arbitrary $\beta E_b$ and different power-law parameters, the inset corresponds to the low $\beta E_b$

In Fig.3, we showed a numerical result for the integral Eq. (32) and compared the result for arbitrary $\beta E_b$ with the Kramers result for $\beta E_b \gg 1$. The Kramers MFPT for the low damping with the power-law distribution is given in the appendix (see Eq.

(A.14)), where these two diffusion coefficients satisfy $D'/D = (2\pi/\omega_0)^2 E_b/m$.

Fig.3 showed a similar trend to Fig.2. The ratio $\tau/\tau_K$ only depends on $\beta E_b$ if $\beta E_b$ is small and it becomes larger with the increase of $\beta E_b$. For $\kappa = 0.04$, the maximum is $\tau/\tau_K = 1.27$ at $\beta E_b = 3.80$, which is also smaller than that in the case of $\kappa = 0$ (i.e. $\tau/\tau_K = 1.30$ and $\beta E_b = 4.17$). Then the ratio reduces gradually. For instance, $\tau/\tau_K = 1.10$ at $\beta E_b = 8.2$ and $\tau/\tau_K = 1.07$ at $\beta E_b = 10$, all of which are much closer to 1 than in the case when $\kappa = 0$. Therefore, the Kramers low damping result with the power-law distribution is approximately accurate for a relatively low $\beta E_b$ if the diffusion coefficient is constant.

## 4. Josephson junction in a finite barrier regime

A Josephson junction consists of two superconductors coupled by a weak link and it has received much attention in both theory and experiment [3, 4, 41, 44, 45]. At temperatures sufficiently close to the transition temperature, thermal fluctuations can disrupt the coupling of the phases of the order parameters of two superconductors separated by a thin insulating barrier. The Josephson current thereby acquires a noise voltage with a nonzero average value [44]. The electrodynamics of a Josephson junction is typically described in terms of an equivalent current-biased circuit consisting of a capacitor (with capacitance $C$) and a resistor (with resistance $R$). In this model, the junction equation of motion is approximately a Langevin-like equation of the form [45],

$$\frac{d^2\theta}{dt^2} = -\frac{dU}{d\theta} - \eta \frac{d\theta}{dt} + \varsigma(t), \qquad (33a)$$

$$\langle \varsigma(t) \rangle = 0, \quad \langle \varsigma(t)\varsigma(t') \rangle = 4\eta \omega_j^2 / \gamma_0 \, \delta(t-t'), \qquad (33b)$$

where $\theta$ is the phases of the order parameters, $U(\theta)$ is the potential function, $U(\theta) = \omega_j^2 (\alpha\theta + \cos\theta)$, $\eta$ is the friction parameter, $\eta = 1/RC$, $\omega_j$ is the plasma frequency, $\omega_j = (2\pi I_C / \Phi_0 C)^{1/2}$, $\gamma_0$ is the dimensionless temperature parameter, $\gamma_0 = \Phi_0 I_C / \pi k_B T$, $\Phi_0$ is the magnetic flux quantum, $\alpha \equiv I_0 / I_C$, $I_0$ is the bias current of the circuit, and $I_C$ is the critical current. Generation of a noise voltage with a nonzero average value can be considered as a Brownian particle performing its motion in a potential energy $U(\theta)$ with the damping coefficient $\eta$ and the random force $\varsigma(t)$ escapes from a metastable state, so it can be directly treated by Kramers escape theory. Since the escape process is stochastic and the bias current of each escape is also stochastic, which result in the distribution function $f(I_0)$ measured in the experiments. According to the relationship between the distribution function (here $f(I_0)$ is replaced by $f(\alpha)$) and MFPT [41],

$$f(\alpha) = \left(\frac{d\alpha}{dt}\right)^{-1} \tau^{-1} \exp\left[-\int_0^\alpha \left(\frac{d\alpha'}{dt}\right)^{-1} \tau^{-1} [\beta E_b(\alpha')] d\alpha'\right], \tag{34}$$

the MFPT can be indirectly obtained and testified.

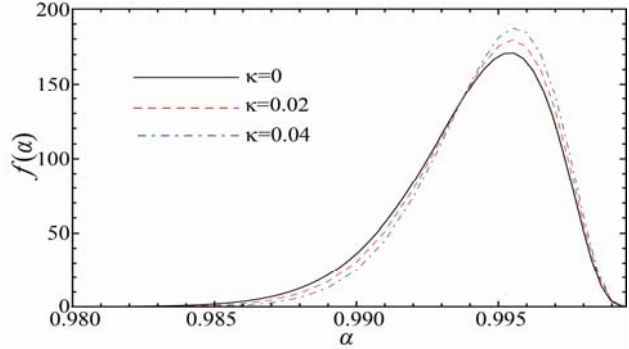

Fig.4 Distribution function in finite barrier regime for different power-law parameters

In section 3 we derived an expression of the MFPT for the finite $\beta E_b$ under the standard FDR and generalized FDR, respectively. Now, we apply them to the distribution function $f(\alpha)$ and study how the power-law parameter $\kappa$ affects $f(\alpha)$ in the finite barrier regime. In Fig.4, we plot $f(\alpha)$ for arbitrary $\alpha$ with $\kappa = 0$, $\kappa = 0.02$ and $\kappa = 0.04$, other parameters in the realistic junction [41] are $T=2K$, $I_C=0.2mA$ and $RCd\alpha/dt=10^{-3}$.

In Fig.4 we show that the power-law parameter $\kappa$ has a significant effect on the distribution function $f(\alpha)$ for about $\alpha$ in the range 0.9840~0.9995, but don't have much effect on $f(\alpha)$ for about $\alpha < 0.9800$) and about $\alpha > 0.9995$, where $\alpha = 0.9800$ and 0.9995 correspond to $\beta E_b$=12.7, and 0.05 ($\beta E_b = \frac{\gamma_0}{2}\left[-\pi\alpha + 2\left(\alpha\sin^{-1}\alpha + \sqrt{1-\alpha^2}\right)\right]$) , respectively. For the lower $\beta E_b$, as long as particles in the well have small energy, they can pass the low barrier quickly regardless of whether the system is in equilibrium or not. For higher $\beta E_b$, the particles which can overcome the higher barrier are quite rare, so the probability of crossing is rather small, and this process corresponds to the zero-voltage state. In the transition regime of $\beta E_b$, some particles gain enough energy, leaves the well and starts running down to the potential slope with an average velocity, and this process corresponds to a nonzero-voltage state [45]. If there is no noise, the system will make a deterministic transition from the zero-voltage state to the nonzero-voltage state at $I=I_C$ which corresponds to. $\beta E_b$=0 However, if there is noise, it can cause the escape from the potential well even if a barrier is present ($\beta E_b \neq 0$). Therefore, noise produces the transition regime of $\beta E_b$. It is known that physically the power-law parameter $\kappa$ reflects the influence of the environment on the system. The external interaction can strengthen or weaken noise, which leads to more escapes ($\kappa<0$) or fewer escapes ($\kappa>0$). So, $\kappa>0$ results in fewer escapes for particles and these escapes occur when the current $I$ is closer to $I_C$ compared to $\kappa=0$, which leads to a narrower distribution. This situation is analogous to superdiffusive effect and subdiffusive effect in nonlinear diffusive media [46] and the physical meaning of power-law parameter $\kappa$ may be explained as an anomalous escape factor.

**5. Conclusion**

According to the MFPT theory, we have derived the expression of the MFPT in an energy-diffusion controlled regime with a power-law distribution. In general, the MFPT theory is established with the assumption that the system follows a thermal equilibrium distribution. In fact, actual systems are usually complex, open, and not in equilibrium; hence the MFPT theory needs to be generalized to nonequilibrium systems. It is considered that a system far away from equilibrium does not have to relax to a thermal equilibrium state with a MB distribution, but might asymptotically approach a nonequilibrium stationary state with a power-law distribution. Therefore, the MFPT theory can be established under the condition of a power-law distribution. We have compared the finite barrier result with the Kramers infinite barrier result in the case of the power-law distribution and MB distribution, respectively. It is shown that the MFPT with the power-law distribution extends the Kramers' low damping result to relatively low $\beta E_b$. We applied our generalized MFPT result to a Josephson junction, and study how the power-law parameter $\kappa$ affects the escape behavior of particles and the current distribution function in the experiment of Josephson junctions. Since $\kappa \neq 0$ measures a distance away from thermal equilibrium and reflects a influence of the environment on the system, $\kappa<0$ and $\kappa>0$ respectively stand for stronger and weaker influence from the environment, which are analogous to superdiffusive effect and subdiffusive effect in nonlinear diffusive media, so $\kappa$ may be considered as an anomalous escape factor. Furthermore, it will be expected that our generalized MFPT is a more general and desired result for describing nonequilibrium systems.

**Appendix**

In the energy region, Fokker-Planck equation can be written [3] as

$$\frac{\partial \rho}{\partial t} = \frac{\omega(I)}{2\pi} \frac{\partial}{\partial E}(\gamma I \rho) + \frac{\omega(I)}{2\pi} \frac{\partial}{\partial E}\left(D' \frac{\omega(I)}{2\pi} \frac{\partial \rho}{\partial E}\right), \tag{A.1}$$

and the stationary state solution is solved exactly,

$$\rho_s \sim \exp\left(-\int \frac{2\pi \gamma I}{\omega D'} dE\right). \tag{A.2}$$

It is well known that, when a system reaches its thermal equilibrium state, there exists a standard fluctuation-dissipation relation (FDR) [3], i.e. $D' = 2\pi \gamma I / \omega \beta$, and Eq. (A.2) follows MB distribution $\rho_s \sim \exp(-\beta E)$. However, the coupling to the bath is very weak and the time to reach thermal equilibrium very long in low damping systems, so the particles may escape before thermal equilibrium and thus the standard

FDR needs to be generalized to the systems away from equilibrium. We can follow Du's line of derivation in Ref. [30] and obtain the generalized FDR,

$$D' = \frac{2\pi}{\omega} \gamma I \beta^{-1} (1 - \kappa \beta E). \tag{A.3}$$

In energy space, the continuity equation [3] is

$$\frac{\partial \rho}{\partial t} = -\frac{\omega(I)}{2\pi} \frac{\partial J}{\partial E}. \tag{A.4}$$

Substitute Eq. (A.4) into Eq. (A.1) and the current $J$ is,

$$J = -\frac{D'\omega(I)}{2\pi} \exp\left(-\int \frac{2\pi I \gamma}{D'\omega(I)} dE\right) \frac{\partial}{\partial E}\left(\rho \cdot \exp\left(\int \frac{2\pi I \gamma}{D'\omega(I)} dE\right)\right). \tag{A.5}$$

Supposing that the distribution function can be written in the form, $\rho(E) = \xi(E)\rho_s(E)$, then Eq. (A.5) becomes

$$J = -\frac{D'\omega(I)}{2\pi} \rho_s(E) \frac{\partial \xi(E)}{\partial E}. \tag{A.6}$$

Integrating over $E$ on both sides of Eq. (A.6) ($J$ is treated as a constant), we have

$$J = \frac{\xi(E) - \xi(E_b)}{\int_{k_B T}^{E_b} \frac{2\pi}{D'\omega(I)} \rho_s^{-1} dE}. \tag{A.7}$$

An absorbing boundary condition in which the particles are absorbed once they arrive at the boundary are used, i.e. $\rho_b(E) = 0$, $\xi(E_b) = 0$. In the bottom of the well, the steady-state distribution is maintained so $\xi(E) = 1$, therefore Eq. (A.7) is simplified as

$$J = 1 \bigg/ \int_{k_B T}^{E_b} \frac{2\pi}{D'\omega(I)} \rho_s^{-1} dE. \tag{A.8}$$

Assume that the potential energy in the well has a minimum $U_{\min} = 0$, and that it can be diagonal and expanded as a harmonic function near the minima $x_a$, namely, $V(x) = \frac{1}{2} \omega_0^2 (x - x_a)^2$, where $\omega_0$ is the frequency. The population $n$ inside the potential well is then calculated,

$$n = \iint \rho_s(x, p) dx dp = Z^{-1} \frac{2\pi}{\omega_0 \beta} \begin{cases} \frac{1}{-\kappa} \Gamma^2\left(-\frac{1}{\kappa} - \frac{1}{2}\right) \bigg/ \Gamma^2\left(-\frac{1}{\kappa}\right), & (-2 < \kappa < 0) \\ \frac{1}{\kappa} \Gamma^2\left(\frac{1}{\kappa} + 1\right) \bigg/ \Gamma^2\left(\frac{1}{\kappa} + \frac{3}{2}\right), & (\kappa > 0) \end{cases}. \tag{A.9}$$

Hence, in the low damping the escape rate with the power-law distribution is

$$k = \frac{J}{n} = \frac{1}{n\int_{k_BT}^{E_b} \frac{2\pi}{D\omega(I)} \rho_s^{-1} dE} = \frac{1}{nZ\int_{k_BT}^{E_b} \frac{2\pi}{D\omega(I)} (1-\kappa\beta E)_+^{-1/\kappa} dE}. \tag{A.10}$$

Put Eq. (A.3) into Eq. (A.10) to simplify the integration of the denominator. Two possible cases are taken into consideration.

(I). The friction coefficient $\gamma$ is a constant, but the diffusion coefficient is a function of the energy, i.e. $D(E) = \frac{2\pi}{\omega} \gamma I \beta^{-1} (1-\kappa\beta E)$. When the barrier height $E_b$ is large relatively to $k_BT$, the main contribution to this integral comes from the ratio of $E_b/k_BT$, so that we may take $I$ to have the value $I_b$ corresponding to the energy trajectory through the saddle point $b$ [3]. Thus, the integral in Eq. (A.10) may be approximately written as

$$\int_{k_BT}^{E_b} \frac{\beta}{I\gamma} (1-\kappa\beta E)^{-\frac{\kappa+1}{\kappa}} dE \approx \frac{(1-\kappa\beta E_b)^{-\frac{1}{\kappa}}}{I_b\gamma}. \tag{A.11}$$

Then the escape rate $k$ is

$$k = \frac{\omega_0 I_b \gamma}{2\pi} (1-\kappa\beta E_b)^{\frac{1}{\kappa}} \begin{cases} -\kappa\beta \Gamma^2\left(-\frac{1}{\kappa}\right) / \Gamma^2\left(\frac{1}{\kappa}-\frac{1}{2}\right), & (-2<\kappa<0) \\ \kappa\beta \Gamma^2\left(\frac{1}{\kappa}+\frac{3}{2}\right) / \Gamma^2\left(\frac{1}{\kappa}+1\right), & (\kappa>0) \end{cases}. \tag{A.12}$$

In the limit $\kappa \to 0$, this becomes the traditional Kramers escape rate for the low damping, $k = \frac{\omega_0 \beta I_b \gamma}{2\pi} e^{-\beta E_b}$.

(II). The diffusion coefficient $D$ is a constant, and the frequency can be approximately taken as $\omega(I) \approx \omega_0$. Eq.(A.10) is written as

$$\int_{k_BT}^{E_b} \frac{2\pi}{D\omega(I)} (1-\kappa\beta E)_+^{-1/\kappa} dE \approx \frac{2\pi}{D\omega_0} \int_{k_BT}^{E_b} (1-\kappa\beta E)^{-\frac{1}{\kappa}} dE = \frac{2\pi}{D\omega_0} \frac{(1-\kappa\beta E_b)^{\frac{\kappa-1}{\kappa}}}{\beta(1-\kappa)}. \tag{A.13}$$

And then escape rate $k$ is

$$k = \left(\frac{\omega_0}{2\pi}\right)^2 \beta^2 D'(1-\kappa)(1-\kappa\beta E_b)^{\frac{1}{\kappa}-1} \begin{cases} -\kappa\Gamma^2\left(-\frac{1}{\kappa}\right)/\Gamma^2\left(-\frac{1}{\kappa}-\frac{1}{2}\right), & (-2<\kappa<0) \\ \kappa\Gamma^2\left(\frac{1}{\kappa}+\frac{3}{2}\right)/\Gamma^2\left(\frac{1}{\kappa}+1\right), & (\kappa>0) \end{cases}, \tag{A.14}$$

where diffusion coefficient $D'$ takes $D' = \frac{2\pi}{\omega_0} \gamma_b I_b \beta^{-1} (1-\kappa\beta E_b)$. In the limit $\kappa \to 0$, Eq. (A.14) becomes the traditional Kramers low damping result, i.e. $k = \frac{1}{2\pi} \omega_0 \beta I_b \gamma_b e^{-\beta E_b}$. Thus, we have generalized the Kramers escape rate for the low damping to the systems with the power-law distribution.

**Acknowledgments**

This work is supported by the National Natural Science Foundation of China under grant No 11175128 and by the Higher School Specialized Research Fund for Doctoral Program under grant No 20110032110058.